\documentclass[preprint,aps,prb,superscriptaddress]{revtex4}
\usepackage{multirow}
\usepackage{graphicx}
\usepackage{dcolumn}
\usepackage{bm}

\usepackage{titlesec}
\titleformat*{\section}{\flushleft \bf \large}
\titleformat*{\subsection}{\flushleft \bf}
\titleformat*{\subsubsection}{\flushleft}
\begin{document}

\title{Dissociation of O$_2$ molecules on strained Pb(111) surfaces}
\author{Yu Yang}
\affiliation{LCP, Institute of Applied Physics and Computational
Mathematics, P.O. Box 8009, Beijing 100088, People's Republic of
China}
\author{Ping Zhang}
\thanks{Corresponding author. zhang\_ping@iapcm.ac.cn}
\affiliation{LCP, Institute of Applied Physics and Computational
Mathematics, P.O. Box 8009, Beijing 100088, People's Republic of
China} \affiliation{Center for Applied Physics and Technology,
Peking University, Beijing 100871, People's Republic of China}

\begin{abstract}
By performing first-principles molecular dynamics calculations, we
systematically simulate the adsorption behavior of oxygen molecules
on the clean and strained Pb(111) surfaces. The obtained molecular
adsorption precursor state, and the activated dissociation process
for oxygen molecules on the clean Pb surface are in good agreements
with our previous static calculations, and perfectly explains
previous experimental observations [Proc. Natl. Acad. Sci. U.S.A.
104, 9204 (2007)]. In addition, we also study the influences of
surface strain on the dissociation behaviors of O$_2$ molecules. It
is found that on the compressed Pb(111) surfaces with a strain value
of larger than 0.02, O$_2$ molecules will not dissociate at all. And
on the stretched Pb(111) surfaces, O$_2$ molecules become easier to
approach, and the adsorption energy of the dissociated oxygen atoms
is larger than that on the clean Pb surface.

\end{abstract}

\maketitle

\section{Introduction}

Oxygen molecules in the atmosphere are thermodynamically allowed to
react with most metal surfaces, causing the formation of thin oxide
films \cite{Lawless1974,Kung1989,Henrich1994,Ciacchi04}. This
phenomenon has been widely taken advantage of during the
fabrications of heterogeneous catalysts, gas sensors, dielectrics,
and corrosion inhibitors \cite{Kung1989,Henrich1994}. However,
understanding the oxidation mechanisms from the atomic view, as well
as artificially manipulating the oxide products are still not mature
yet. Although the adsorption and dissociation processes of oxygen
molecules on most metal surfaces have been studied widely and deeply
\cite{Ertl1993,Madix1994,Osterlund1997,Honkala00,Hellman05}, some
more practical problems like the influences of surface morphology on
the oxygen dissociation processes are still lacking attentions. With
the latest development of surface science, experiment researchers
are now able to manipulate surface morphologies at the nanometer
level \cite{Thurmer01,Hong03,Jiang04,Lin05,Chan06,Libuda05}, which
implies that by controlling the surface morphology, one may be able
to manipulate the oxidation processes and the corresponding reaction
products of metal surfaces. Recently, Th\"{u}rmer {\it et al.}
\cite{Thurmer02} and Ma {\it et al.} \cite{Ma07} studied the
manipulation of surface impurity and quantum size effect on the
oxidation of epitaxial Pb(111) crystallites, respectively. The
former revealed that pure Pb(111) crystallites were resistant to
oxidation, but once nucleated by surface impurities, monolayer films
of lead oxide would grow on the Pb(111) surface in an autocatalytic
process \cite{Thurmer02}. The latter revealed that the adsorption
rate of oxygen and oxidation rate of Pb were both modulated by the
quantum size effect of ultrathin Pb(111) films, and showing even-odd
oscillations \cite{Ma07}. Therefore, studying the modulation of
different surface morphologies on the oxidation of metal is very
meaningful, and here in this paper, we study the influences of
surface strain on the oxidation of the Pb(111) surface.

In addition, studying the influences of surface strain on oxidation
is also practically very needed. The above two experimental studies
about surface oxidation of Pb (Refs. \onlinecite{Thurmer02} and
\onlinecite{Ma07}) are both carried out on epitaxially growed
Pb(111) terraces, which are deposited on the Ru(0001) and Si(111)
surfaces respectively \cite{Thurmer02,Ma07}. Since there are 9\% and
5\% lattice mismatches between Pb and Si \cite{Wei02}, Ru, the
deposited Pb(111) terraces might contain some inherent strains. This
fact also motivates our present study. More importantly, most
experimental studies on surface oxidation of metals are also carried
out on epitaxially deposited metal surfaces
\cite{Madix1994,Libuda05}, which inevitably introduces strains on
metal surfaces. In this way, studying the influences of strain on
surface oxidation of metals has a broad significance. In the
following of the paper, we first build the atomistic model for
oxygen dissociation on the Pb(111) surface, based on
first-principles molecular dynamics (FPMD) simulations, and then
study the relationship between surface strain and the dissociation
processes of O$_2$ molecules on differently strained Pb(111)
surfaces.

\section{Calculation method}

Our calculations are performed using the Vienna {\it ab-initio}
simulation package \cite{VASP}. The PW91 \cite{PW91} generalized
gradient approximation and the projector-augmented wave potential
\cite{PAW} are employed to describe the exchange-correlation energy
and the electron-ion interaction, respectively. The cutoff energy
for the plane wave expansion is set to 400 eV. The molecular
dynamics (MD) simulations are performed using the Verlet algorithm
with a time step of 1 fs within the microcanonical ensemble. In our
present study, the Pb(111) surface is modeled by a periodically
repeated slab of four Pb layers separated by a vacuum region
correspondent to six metal layers. We consider a (3$\times$3)
surface unit cell with 9 Pb atoms in each atomic layer. The surface
Brillouin zone is sampled by a 3$\times$3 $k$-point distribution
using the Monkhorst-Pack scheme \cite{Monkhorst}. The calculated
lattice constant of bulk Pb and the bondlength of isolated O$_{2}$
are 5.03 \AA~ and 1.24 \AA, respectively, in good agreements with
the experimental values of 4.95 \AA~\cite{Wyckoff1965} and 1.21
\AA~\cite{Huber1979}. The O$_{2}$ is placed on one side of the slab,
namely on the top surface, whereas the bottom Pb layer is fixed. All
other Pb atoms as well as the oxygen atoms are free to move during
the geometry optimizations and MD simulations.

\section{Results and discussion}

Before simulating the adsorption of O$_2$ molecules, we firstly
optimize the geometry of the clean Pb(111) surface. During the
surface relaxation, the Pb atoms at the outmost layer tend to relax
inward, because of the asymmetry of the electron density at the
surface. In contrast, Pb atoms at the second layer tend to relax
outward \cite{Yang08}. With respect to the interlayer spacing along
the [111] direction of bulk Pb, the first interlayer spacing of the
Pb(111) surface is compressed by 2.26\%, whereas the second
interlayer spacing is expanded by 4.35\%. We also perform geometry
optimizations for the compressively and stretchingly strained
Pb(111) surfaces. The surface strains are simultaneously added on
the $x$ and $y$ directions, with plus and minus values representing
for stretched and compressed surfaces respectively. The Pb atoms are
free to relax during the geometry optimizations of the compressed or
stretched surfaces. From the optimized geometries, we find that
within the surface strains between $\pm$0.1, no obvious structure
transformation happens for the studied Pb(111) surface.

We then perform our FPMD simulations to study oxygen dissociation on
the Pb(111) surface, which are started with different orientations
of an O$_2$ molecule placed over different surface sites. There are
four high symmetry sites on the Pb(111) surface, namely the top (T),
bridge (B), fcc- (FH), and hcp-hollow (HH) sites, and the adsorbed
O$_2$ molecule has three high-symmetry orientations ([0$\bar{1}$1],
[$\bar{2}$11], and [111] directions). Thus in total, we consider 12
high-symmetry trajectories for the adsorption of O$_2$. At the
beginning of all these trajectories, the mass center of the O$_2$
molecule is initially set to be 4 \AA~ away from the metal surface.
The substrate Pb atoms are initially set at rest while the adsorbed
oxygen atoms are initially with a kinetic energy.

Figure 1 shows the free electronic energy evolutions for two FPMD
simulations of an parallel O$_2$ molecule adsorbing at the surface
hcp hollow site of Pb. The initial kinetic energy of the O$_2$
molecule is 0.2 and 0.6 eV respectively. We can see clearly that the
free electronic energy before t=300 fs evolves similarly for the two
O$_2$ molecule with different initial kinetic energies. As we will
discuss in the following, the energy evolution at this period
corresponds to the repulsion of surface electrons to the O$_2$
molecule. After t=300 fs, the two O$_2$ molecules show different
adsorption behaviors. The adsorption system of the O$_2$ molecule
with an initial kinetic energy of 0.6 eV experiences an energy
reduction of about 1.6 eV. As we will see later, this reduction of
energy corresponds to the dissociative adsorption of O$_2$. At
meantime, the adsorption system with the O$_2$ molecule with an
initial kinetic energy of 0.2 eV has few changes in the free
electronic energy. This results clearly indicate that the
dissociative adsorption of O$_2$ molecules on the clean Pb(111)
surface needs to overcome an energy barrier, with the value of
between 0.2 and 0.6 eV.

We have previously studied in depth the molecular and atomic
adsorption of oxygen on the Pb(111) surface, by using static
first-principles calculations \cite{Yang08,Sun08}. It has been
revealed that O$_2$ adsorbs barrierlessly into the molecular
adsorption precursor state on the clean Pb(111) surface. In these
precursor states, the O$_2$ molecule adsorbs at surface hcp or fcc
hollow sites, with no spin any more because of the electronic
hybridizations with surface electrons, and the precursor state at
surface hcp hollow site is the most stable one \cite{Yang08}. In our
recent studies, we find that the dissociation energy barrier for
O$_2$ from the molecularly adsorbed precursor states is much smaller
than that directly from the gas-like molecules, and the minimum
energy dissociation path is from the most stable precursor
adsorption state of O$_2$ on the Pb(111) surface \cite{Hu11}. Here
through FPMD simulations, we find that the O$_2$ molecule along the
top and bridge trajectories will firstly moves to the hcp or fcc
hollow sites before its adsorption, and the energy evolutions along
the hcp and fcc trajectories are very similar to each other.
Therefore, we will mainly discuss the results of the hcp-hollow
trajectory at subsequent discussions, which already contains most of
the related information during the adsorption and dissociation of
O$_2$ molecules.

At another side, although the ground electronic state of an O$_2$
molecule is the spin-triplet state, the adsorption system in both
the molecular adsorption precursor state \cite{Yang08} and the
atomic adsorption states are found to be nonmagnetic \cite{Sun08}.
These results suggest that spin polarization has little importance
to the adsorption behavior of O$_2$ on the Pb(111) surface. Our test
FPMD calculation also proves this result, as the spin of O$_2$
quickly quenches to be zero when approaching the Pb(111) surface.
Therefore, to simplify the calculations and discussions, our FPMD
calculations are all spin-nonpolarized.

Our ensemble for the adsorption system of O$_2$ on the Pb(111)
surface is a microcanonical one, thus the total energy (i.e.
summation of free electronic energy, Madelung energy of ions, and
kinetic energy of ions) is conserved. This fact is clearly shown in
Fig. 2(a), where the free electronic energy of the adsorption system
at t=0 fs is set to zero. The obtained free electronic energy of the
adsorption system during the adsorption process of an O$_2$ molecule
at the surface hcp hollow site is shown in Fig. 2(b). Figures 2(c)
and 2(d) show the molecule bond length ($d_{\rm O-O}$) and the
height of the O$_2$ center of mass ($h_{\rm O_2}$), respectively. We
can see from Fig. 2(b) that when the O$_2$ molecule approaches the
Pb(111) surface, the free electronic energy of the adsorption system
goes down. This result is in agreement with our previous static
calculations finding that O$_2$ molecule enters into the molecular
adsorption state without any energy barriers \cite{Yang08}, and
accords well with the previous experimental observation that O$_2$
molecules adsorb on the Pb(111) surface at very low temperatures
\cite{Ma07}. The energy reduction when approaching the metal surface
is different from that has been found for the O$_2$/Be(0001) system,
where the free electronic energy of the adsorption system goes up
when O$_2$ approaches the Be surface \cite{Yang10}. As shown in Fig.
2(b), when the O$_2$ molecule approaches too close to the Pb(111)
surface (after t1=90 fs), the free electronic energy goes up
sharply, because of the electronic repulsion from the surface
electrons of Pb to the O$_2$ molecule. And the free electronic
energy of the adsorption system reaches a local maximum at t2=107
fs. From t2=107 fs to t3=127 fs, the O$_2$ molecule moves back from
the Pb(111) surface, and the free electronic energy reduces again.
The geometries of the adsorption system at t1, t2, and t3 are shown
in Figs. 3(a)-3(c). We can see from both Fig. 2(c) and Fig. 3 that
the molecular length does not change much from t1 to t3.

After being repulsed by the Pb(111) surface, the O$_2$ molecule
suspends on the Pb(111) surface. Until t4=551 fs, it enters into the
molecular adsorption state. In the precursor state, the molecular
bond length is enlarged to be 1.43 \AA, indicating that the adsorbed
O$_2$ molecule is superoxide-like \cite{Panas1989,Nakatsuji1993}. It
is because that in the molecular adsorption precursor state, a
certain number of electrons transfer from bonding orbitals to Pb,
while more electrons transfer back from Pb to the antibonding
orbitals of O$_2$ \cite{Yang08}. After climbing an energy barrier of
0.31 eV (from t4=551 fs to t5=601 fs), the O$_2$ molecule then
dissociates into two adsorbed oxygen atoms. As shown in Fig. 2(b),
the atomic adsorption of the two oxygen atoms causes the free
electronic energy of the adsorption system to reduce 2.29 eV at
t6=684 fs. The geometries of the adsorption system from the
molecular adsorption precursor state to the atomic adsorption final
state are shown in Figs. 3(d)-3(f). We can see from Fig. 3(f) that
the final state of the dissociation process is the adsorption state
of two oxygen atoms in two neighboring hollow sites. These results
are in agreements with our recent static calculations \cite{Hu11}.

After dissociation, the distance between the two oxygen atoms is
rapidly enlarged by 3 \AA~ within 150 fs, as shown in Fig. 2(c),
which is very similar to the dissociated oxygen atoms on the Al(111)
surface, where the fast moving oxygen atoms are called as ``hot''
oxygen atoms, and the dissociation mechanism is described into a
``Hot-Atom'' picture \cite{Ciacchi04}. This result indicates that
the oxidation mechanisms of Al and Pb are similar. After $t$=800 fs,
the O$_2$/Pb(111) adsorption system begins to vibrate in its
intrinsic frequencies.

The molecular height $h_{\rm O_2}$ of O$_2$ after dissociation (i.e.
after t=700 fs) is found to fluctuate around 1.25 \AA, indicating
that the oxygen atoms do not penetrate into the Pb(111) surface, and
the underlying Pb layers hardly take part in the interactions with
oxygen. We can also see from Fig. 3 that during the dissociation
process, the topmost Pb atomic layer is distorted. However, the
second Pb layer almost does not change at all, with a negligible
distortion. So, the interaction between O$_2$ molecules and the
Pb(111) surface is very localized on the surface. This result is
also in agreement with our previous difference charge density
analysis for the molecular adsorption of O$_2$ on the Pb(111)
surface \cite{Yang08}.

The evolution of the electronic structure during the adsorption and
dissociation of O$_2$ molecules on metal surfaces is of great
theoretical importance and thus has been widely studied
\cite{Behler05,Yang10}. We here also calculate and analyze the
projected density of states (PDOS) of the O$_2$ molecule during its
dissociation process on the unstrained Pb(111) surface. The PDOS of
the O$_2$ molecule in the molecular adsorption, transition, and
atomic adsorption states are shown in Figs. 4(a)-(c) respectively.
As shown in Fig. 4(a), there are four peaks in the $s$- and $p$-PDOS
of the molecularly adsorbed O$_2$, corresponding to the four
molecular orbitals, $\sigma_{2s}$, $\sigma^*_{2s}$, $\sigma_{2p}$,
and $\pi_{2p}$ from low to high energies respectively. The
half-filling $\pi^*_{2p}$ molecular orbital of O$_2$ is broadened
into several energy bands through electronic hybridizations with $p$
electrons of Pb \cite{Yang08}. At t=601 fs, the O$_2$ molecule
evolves into the transition state, with its $\pi_{2p}$ molecular
orbital further vanished through hybridizations with $p$ electrons
of Pb, as shown in Fig. 4(b). In the transition state, because the
two oxygen atoms are not far away from each other, then the
$\sigma_{2s}$, $\sigma^*_{2s}$, and $\sigma_{2p}$ molecular orbitals
still keep their localization characters. After dissociation, all
the molecular orbitals of O$_2$ are broken. The s1 and s2 states
shown in the $s$-PDOS of oxygen in Fig. 4(c) distribute around the
two oxygen atoms, representing for their atomic orbitals. The energy
levels are different for the two oxygen $s$ orbitals because the two
oxygen atoms are at different surface sites [as shown in Fig. 3(f)].
We can see from the PDOS evolution that only the $p$ states of Pb
and oxygen atoms take part in the electronic hybridizations, during
the dissociation process.

By performing a series of FPMD simulations, we find that compressive
and stretching strains influence the oxygen dissociation on the
Pb(111) surface in quite different ways. Once the Pb(111) surface is
compressed over 2\%, we find that oxygen molecules will not
dissociate at all on the Pb(111) surface. As clearly shown in Fig.
5(a), the free electronic energy of the adsorption system for the
compressed Pb(111) surface does not reduce after the surface
repulsion to the adsorbing O$_2$ molecule, indicating that no
chemisorption happens. In contrast, O$_2$ molecules can also
dissociate on the stretched Pb(111) surfaces. We can see from Fig.
5(a) that the free electronic energy of the adsorption system for
the stretched Pb(111) surface reduces by about 2.65 eV after t=500
fs, because of the dissociative chemisorption of O$_2$. After that,
the distance between the dissociated oxygen atoms on the 2\%
stretched Pb(111) surface is rapidly enlarged to be 4 \AA~ within
100 fs, as shown in Fig. 5(b). This result indicates that stretching
strain does not change the ``Hot-Atom'' dissociation mechanism of
O$_2$ molecules on the Pb(111) surface. The molecular height of
O$_2$ from the Pb surface ($h_{\rm O_2}$) on the stretched surface,
as shown in Fig. 5(c), is always smaller than that on the clean
Pb(111) surface, indicating that the chemisorbed oxygen atoms are
nearer to the stretched surface than to the unstrained Pb(111)
surface.

As we have discussed, charge transfer from the Pb surface to the
antibonding orbital of the O$_2$ molecule is very important for the
adsorption interaction between them. Thus we can understand the
above results from the charge density view. On the compressed Pb
surface, surface electrons of Pb are more dense and so their
repulsion to the O$_2$ molecule becomes stronger. This explains why
the O$_2$ molecule does not dissociate on the 2\% compressed Pb
surface. As a proof to the stronger repulsion, the lowest molecular
height during the repulsion process on the compressed Pb surface is
larger than on the unstrained and stretched Pb surfaces, as shown in
Fig. 5(c). On the stretched Pb surface, surface electrons become
looser, and thus become easier to transfer into the antibonding
O$_2$ orbital. This causes that the repulsion force to the O$_2$
molecule is smaller from the stretched Pb surface than from the
unstrained Pb surface, and the dissociated oxygen atoms bond
stronger with Pb atoms of the stretched surface than with that of
the unstrained surface. The smaller molecular height on the
stretched Pb surface also confirms our theory.

To make the influences of stretching strain more clear, we simulate
the adsorption process of O$_2$ molecules on different stretching Pb
surfaces. The free electronic energies correlated with the repulsion
process, i.e., $E{\rm t1}$, $E{\rm t2}$, and $E{\rm t3}$ can be
easily read from the free electronic energy evolutions of FPMD
simulations. The equilibrium energy of the adsorption system after
atomic adsorption of oxygen ($E_{\rm finale}$) can be obtained by
averaging the free electronic energy in a time period of 200 fs.
Based on these quantities, we can define $E_{\rm rep}$=$E{\rm
(t2)}-E{\rm (t1)}$ to scale the repulsion strength of the Pb(111)
surface to the impinging O$_2$ molecule, and $E_{\rm ad}{\rm
(atom)}$=$E_{\rm O_2}+E_{\rm Pb~surf}-E_{\rm finale}$ to scale the
binding strength between dissociated oxygen atoms and the Pb(111)
surface, where $E_{\rm O_2}$ and $E_{\rm Pb~surf}$ are the free
electronic energies of an isolated O$_2$ molecule and the stretched
Pb surface, respectively.

The calculated $E_{\rm rep}$ and $E_{\rm ad}{\rm (atom)}$ on
different stretching Pb(111) surfaces are shown in Fig. 6. One can
see that the more stretched the Pb(111) surface is, its repulsion to
O$_2$ molecules becomes weaker, indicating that O$_2$ molecules are
easier to approach the stretched Pb(111) surface. It is because that
surface electrons of Pb become more dilute after stretching, and so
their repulsion becomes weaker. The reduction of surface electron
density also makes the electronic interactions between oxygen and Pb
more easier on the stretched Pb(111) surfaces. And correspondingly,
the binding energy between oxygen atoms and the stretched Pb(111)
surface becomes larger. Therefore, after the Pb(111) surface being
stretched, O$_2$ molecules are easier to approach the Pb surface
before dissociation, and oxygen atoms bind stronger with Pb atoms
after dissociation.

\section{Conclusion}

In summary, we have systematically investigated the dissociative
adsorption processes of O$_2$ molecules on the Pb(111) surface. By
performing FPMD simulations, we find that the O$_2$ dissociation is
an activated type on the Pb(111) surface, and the obtained molecular
adsorption precursor states from the FPMD simulation are in
agreements with with our previous static calculational results.
Besides, we have also observed the repulsion process when the
impinging O$_2$ molecule approaches too close to the Pb(111)
surface. After a series of FPMD simulations for O$_2$ adsorption on
the strained Pb(111) surfaces, we find that O$_2$ molecules will not
dissociate at all on the compressed Pb(111) surfaces, while on the
stretched Pb(111) surfaces, O$_2$ molecules can still dissociate,
and the atomic bindings between the dissociated oxygen atoms and Pb
atoms become stronger. Through the present studies, we expect to
expand people's knowledge on metal surface oxidation, by pointing
out the specific role of surface strains.

\begin{acknowledgments}
This work was supported by the NSFC under Grants No. 10904004 and
No. 90921003.
\end{acknowledgments}

\clearpage

\noindent\textbf{List of captions} \\

\noindent\textbf{Fig.1}~~~ (Color online). The free electronic
energy of the adsorption system in two first-principles molecular
dynamics (FPMD) simulations for the adsorption of O$_2$ molecule on
the Pb(111) surface. The initial kinetic energy of the O$_2$
molecule is 0.2 and 0.6 eV respectively. \\

\noindent\textbf{Fig.2}~~~ (Color online). (a) and (b) Total energy
and free electronic energy of the adsorption system in a FPMD
simulation for the dissociative chemisorption of an O$_2$ molecule
on the clean Pb(111) surface. The free electronic energy of an O$_2$
molecule plus that of the clean Pb(111) surface is set to energy
zero. (c) and (d) Molecular bond length of O$_2$ ($d_{\rm O-O}$) and
height from the Pb(111) surface of the O$_2$ center of mass ($h_{\rm
O_2}$), in the same FPMD simulation. The t1-t6 are 90, 107, 127,
551, 601, and 684 fs, respectively. \\

\noindent\textbf{Fig.3}~~~ (color online). Snapshots from a FPMD
simulation of the dissociative chemisorption of an O$_2$ molecule on
the Pb(111) surface. Only the outermost two Pb layers are shown. Red
balls represent oxygen atoms, while black and grey balls represent
Pb atoms in the first and second atomic layer, respectively. (a)-(c)
The repulsion process when an O$_2$ molecule approaches to the Pb
surface. (d)-(f) The dissociation process for the O$_2$ molecule
from its molecular adsorption precursor state. The molecular bond
lengths ($d_{\rm O-O}$) are 1.42, 1.44, 1.53, 1.43, 1.87, and 2.83
\AA~ in (a)-(f), respectively. \\

\noindent\textbf{Fig.4}~~~ (Color online). The projected density of
states of the oxygen molecule in the molecular adsorption (a),
transition (b) and atomic adsorption states (c) on the unstrained
Pb(111) surface. The Fermi energies are set to zero, and denoted by
the dotted line. \\

\noindent\textbf{Fig.5}~~~ (a) Free electronic energy of the
adsorption system in three FPMD simulations for the dissociative
chemisorption of an O$_2$ molecule on the compressed, clean and
stretched Pb(111) surfaces. The free electronic energy of an O$_2$
molecule plus that of the strained Pb(111) surface is set to energy
zero. (b) and (c) Molecular bond length of O$_2$ ($d_{\rm O-O}$) and
height from the Pb(111) surface of the O$_2$ center of mass ($h_{\rm
O_2}$), in the three FPMD simulations. \\

\noindent\textbf{Fig.6}~~~ The repulsion energy ($E_{\rm rep}$) and
atomic adsorption energy ($E_{\rm ad}{\rm (atom)}$) for oxygen on
the stretched Pb(111) surfaces. \\

\clearpage

\begin{figure}
\includegraphics[width=1.0\textwidth]{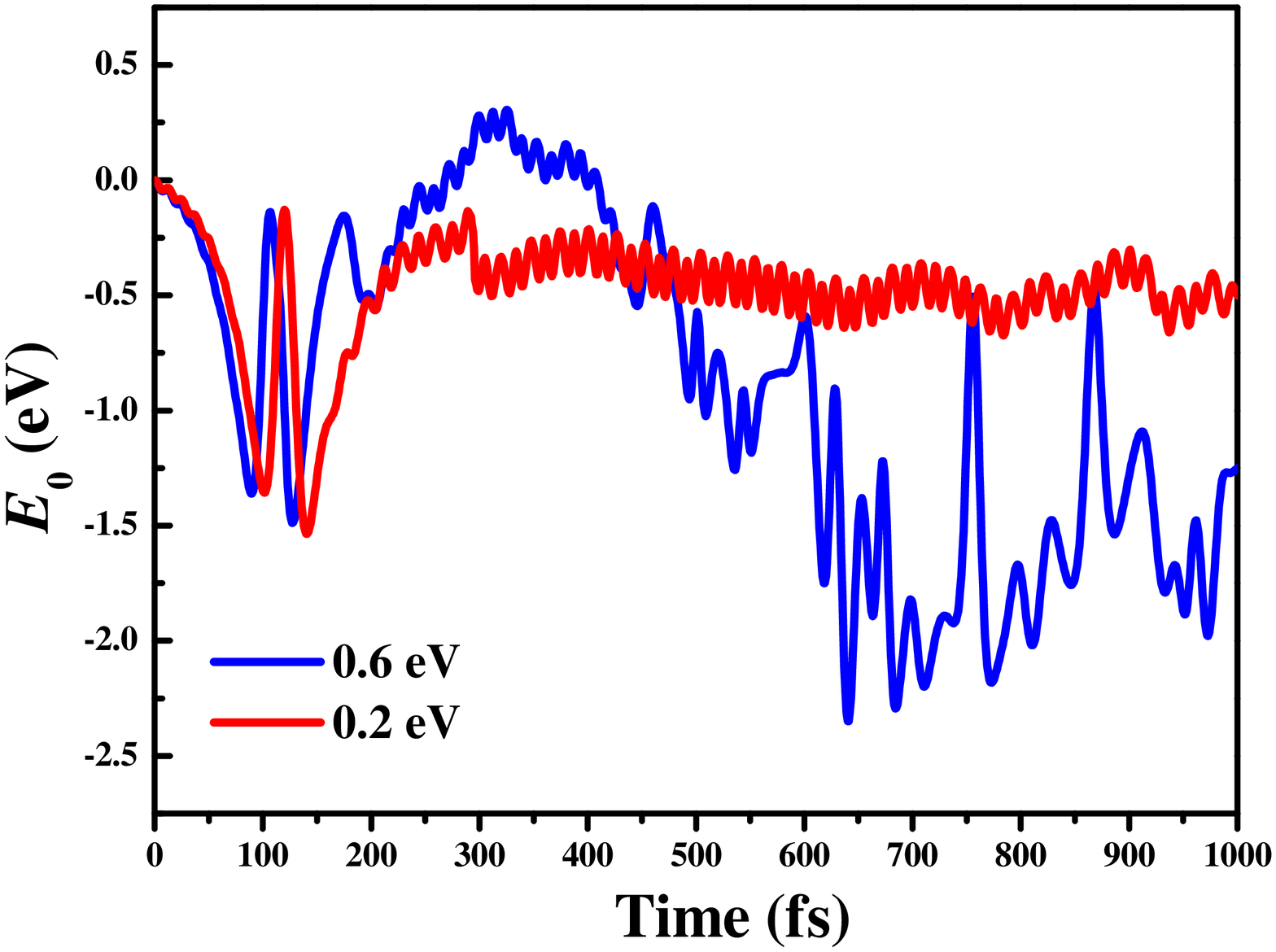}
\caption{\label{fig:fig1}}
\end{figure}
\clearpage
\begin{figure}
\includegraphics[width=1.0\textwidth]{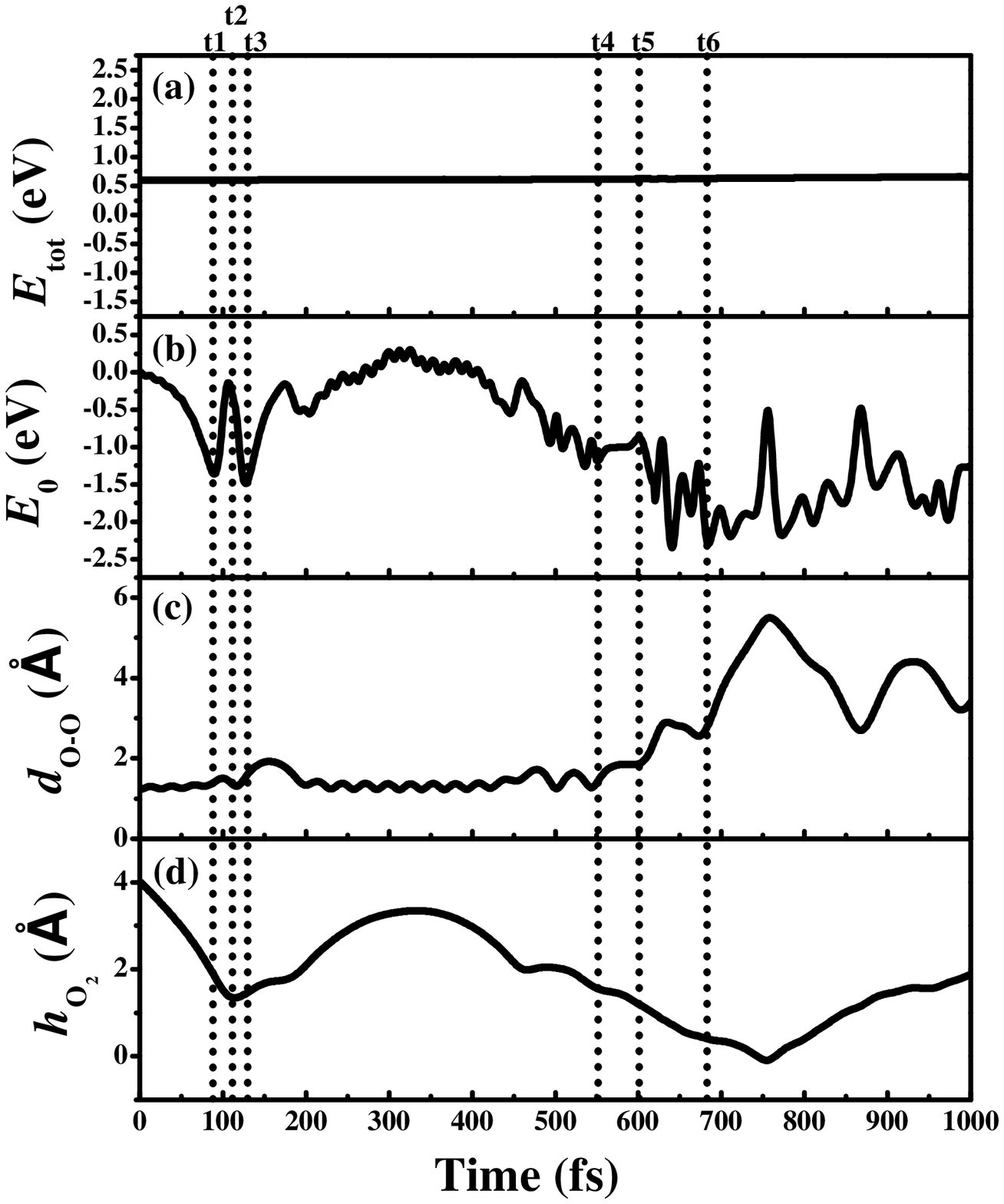}
\caption{\label{fig:fig2}}
\end{figure}
\clearpage
\begin{figure}
\includegraphics[width=1.0\textwidth]{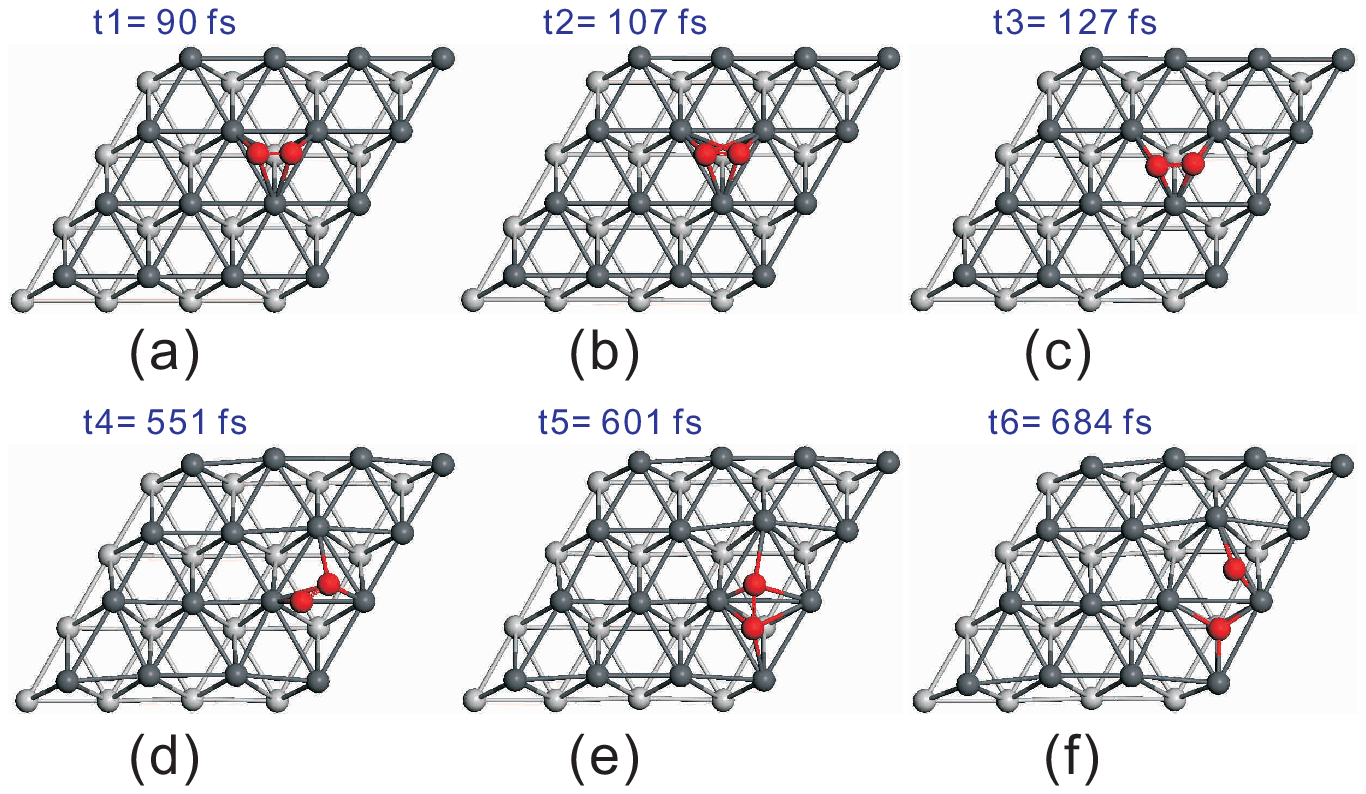}
\caption{\label{fig:fig3}}
\end{figure}
\clearpage
\begin{figure}
\includegraphics[width=1.0\textwidth]{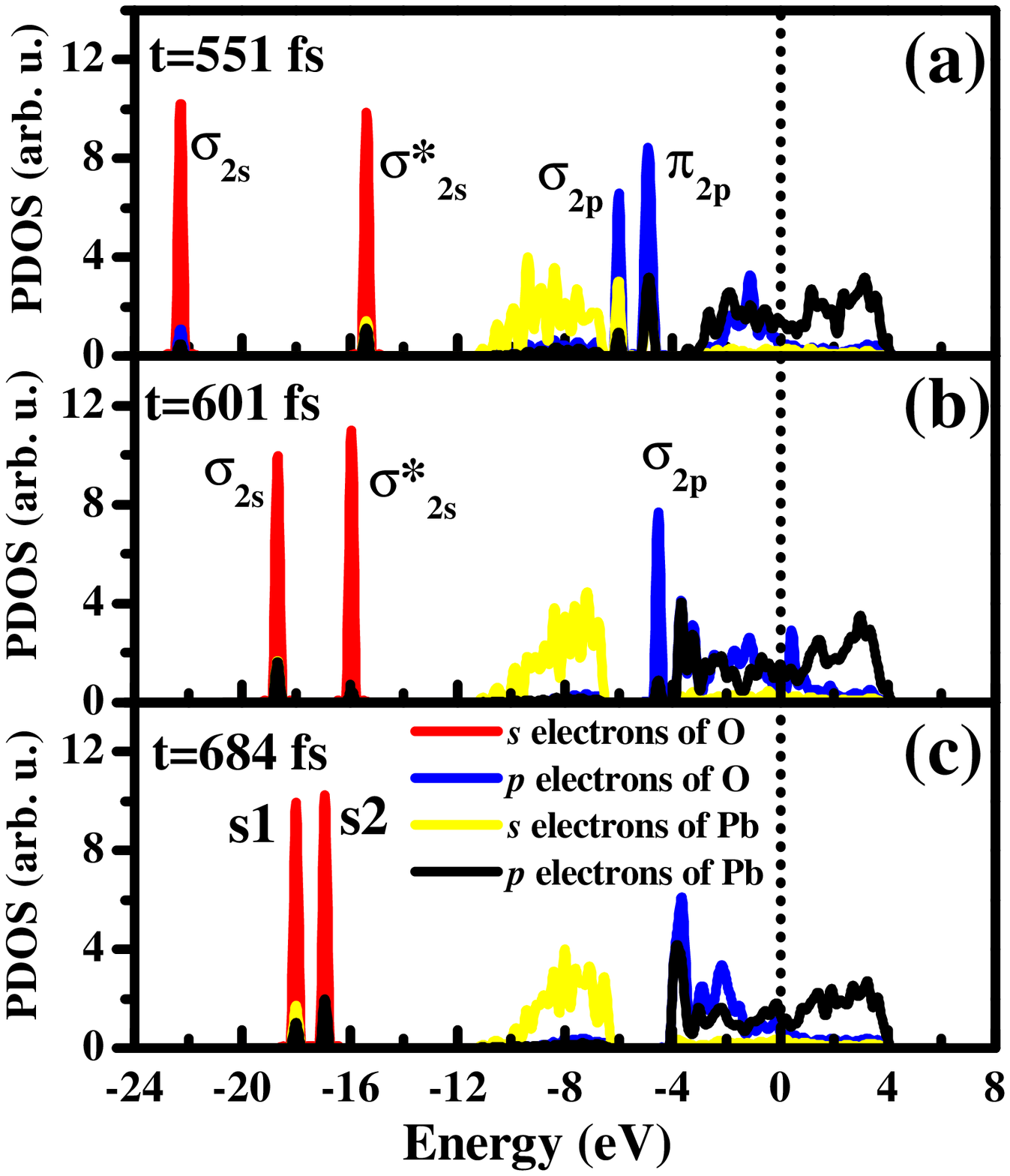}
\caption{\label{fig:fig4}}
\end{figure}
\clearpage
\begin{figure}
\includegraphics[width=1.0\textwidth]{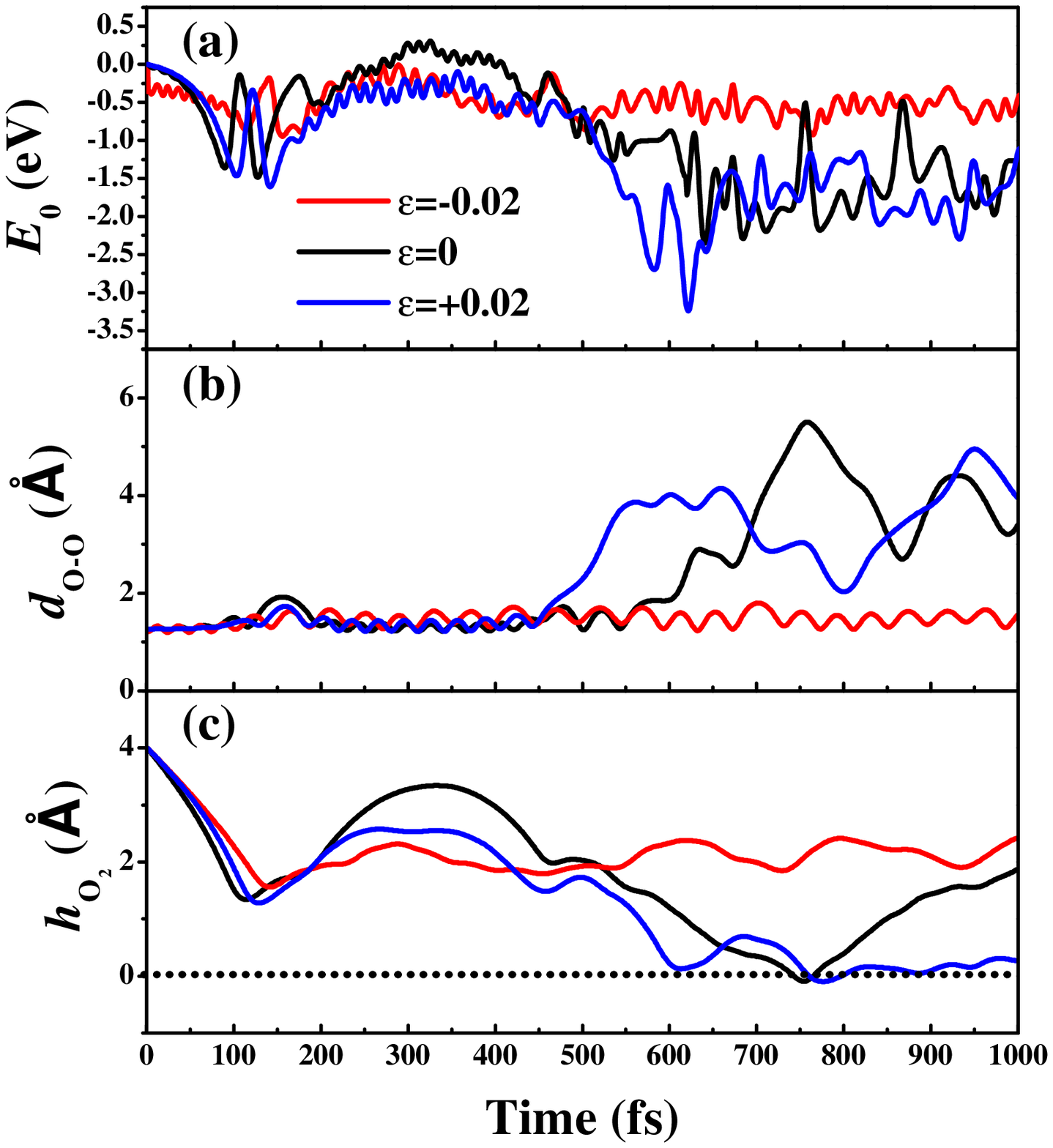}
\caption{\label{fig:fig4}}
\end{figure}
\clearpage
\begin{figure}
\includegraphics[width=1.0\textwidth]{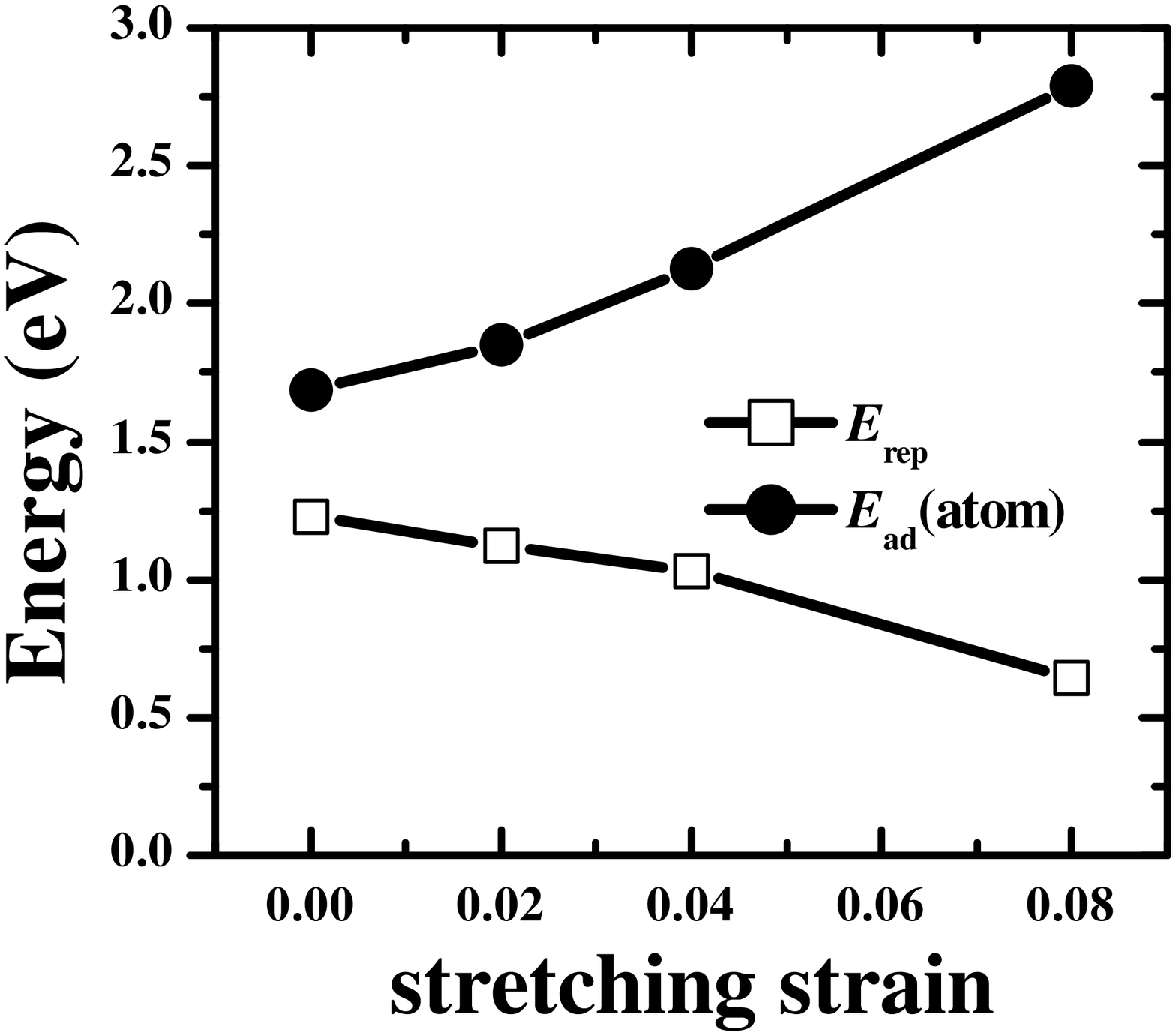}
\caption{\label{fig:fig4}}
\end{figure}
\end{document}